\documentclass[reprint,showpacs,aps,prx,amsmath,amssymb,floatfix,twocolumn,longbibliography]{revtex4-2}
\usepackage{graphicx,epstopdf}
\usepackage{bm}
\usepackage{mathrsfs}
\usepackage{color}
\usepackage[bookmarks,colorlinks,linkcolor=blue,anchorcolor=black,urlcolor=blue,citecolor=blue]{hyperref}
\usepackage{natbib}
\usepackage{CJK}
\usepackage{lipsum}
\usepackage[table]{xcolor}
\usepackage[mathlines]{lineno}

\allowdisplaybreaks[3]

\begin{document}
\begin{CJK*}{UTF8}{gbsn}

\title{Two remote counting events induced by a single photon}
\author{Lida Zhang~(张理达)$^{1,2}$}
\email{zhang\_lida@foxmail.com}
\affiliation{$^{1}$School of Physics, East China University of Science and Technology, Shanghai, 200237, China}
\affiliation{$^{2}$Department of Physics and Astronomy, Aarhus University, Ny Munkegade 120, DK 8000 Aarhus C, Denmark}

\begin{abstract}
Motivated by Einstein's thought experiment that a single quantum particle diffracted after a pinhole could in principle produce an action in two or several places on a hemispherical imaging screen, here we explore theoretically the possibility to simultaneously detect the action of a single photon at two remote places by two respective counting events. This is considered in a cascade quantum system composed of two spatially distant cavities each coupled to a qubit in the ultrastrong coupling regime. We show that a single-photon pulse incident on the two cavities can simultaneously excite the two remote qubits and lead to two subsequent single-photon counting events even when the separation between them is comparable to the spatial length of the photon pulse. Our results does not only justify the aforementioned Einstein's though experiment from a completely new and fundamental perspective but also have practical applications, such as the generation of remote entanglement by a single photon through a dissipative channel which is otherwise unattainable in the strong-coupling regime. 
\end{abstract}

\maketitle
\end{CJK*}

\begin{figure}[t!]\
	\centering\
	\includegraphics[width=0.48\textwidth]{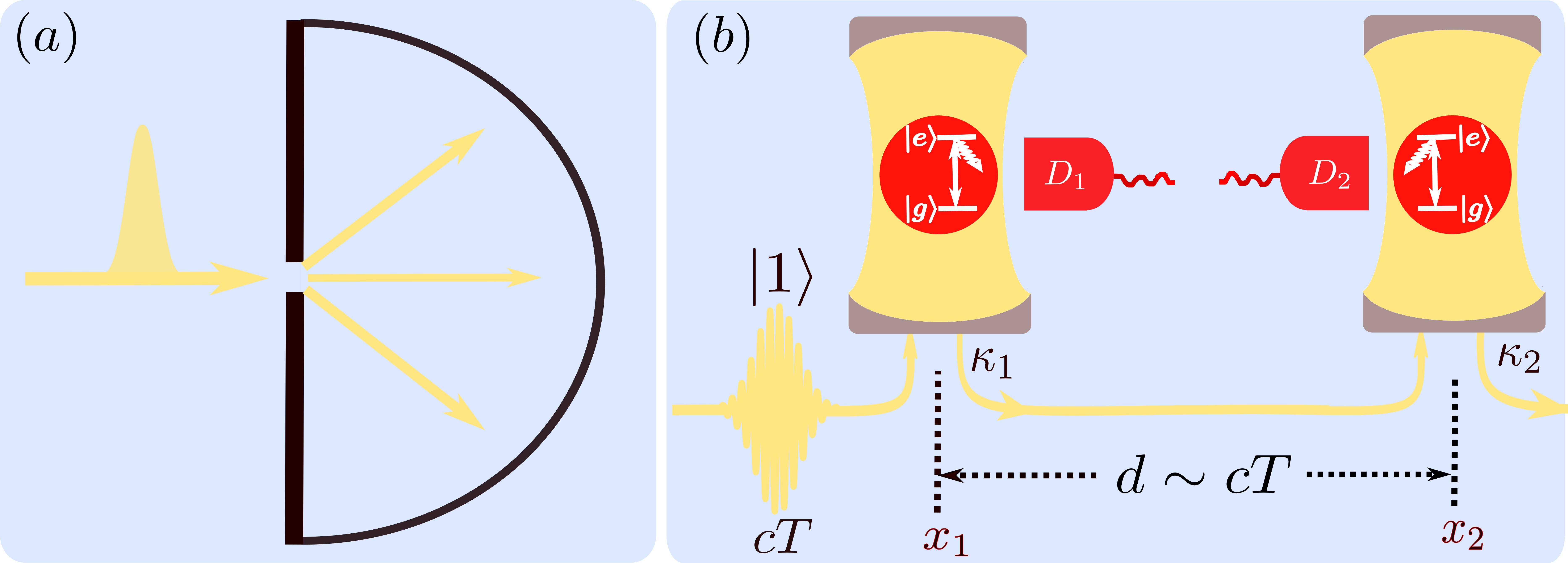}\ 
	\caption{(a) The original thought experiment proposed by Einstein, where a single quantum particle diffracted after a pinhole is then dispersed in space and detected at a hemispherical imaging screen of large radius. Einstein argued that the single quantum particle can produce an action at two or several places. (b) Our  model to demonstrate that a single-photon pulse can simultaneously excite two remote qubits and lead to two subsequent single-photon detection events registered respectively by the detectors $D_{1}$ and $D_{2}$. Each qubit is coupled to a cavity in the USC regime, and the two qubit-cavity subsystems are unidirectionally and dissipatively coupled from the left to the right via the cavity decay $\kappa_{j}$. The separation between the two subsystems $d$ is considered comparable to the spatial length of the incident single-photon pulse $cT$ with $T$ and $c$ being its temporal duration and the speed of light respectively.}\
	\label{fig1}\
\end{figure}

As a prominent critic, Einstein's inquiries into quantum mechanics have played a significant role in advancing the field of quantum physics and fostering innovative concepts. For instance, his questioning on the completeness of quantum mechanics~\cite{Einstein1935PR}, as demonstrated in the Einstein-Podolsky-Rosen~(EPR) paradox, has lead to the concept of quantum entanglement which is vital for quantum information.
Another of Einstein's questions concerns about the interpretation of the wave function, which is the central object of quantum mechanics~\cite{Bacciagaluppi2006arXiv,Ballentine1972AJP}. 
He proposed the thought experiment that a single quantum particle diffracted after a pinhole is then detected at a hemispherical imaging screen of large radius~(see Fig.~\ref{fig1}(a)), and argued that ``if $|\psi|^{2}$ ($\psi$ is the wave function) were simply regarded as the probability that at a certain point a given particle is found at a given time, it could happen that the same elementary process produces an action in two or several places on the screen...'' as quoted from Ref.~\cite{Bacciagaluppi2006arXiv}. 
Inspired by Einstein's thought experiment, here we would like to ask a simple and experimentally verifiable question whether or not a single spatially-extended quantum particle can lead to a simultaneous action at two remote places that can be subsequently detected by two corresponding counting counts. For brevity, this will be referred to two remote counting events induced by a single particle (TRESP) in the following.

\

Though there is no fundamental reason forbidding TRESP, it is conceptually nontrivial since one might naturally expect that a single particle can only trigger a single detection count at one place. Here we propose a simple model to illustrate how a higher-frequency photon pulse propagating in free space can excite nonlocally two remote lower-transition-frequency qubits, which are separated by a distance comparable to the spatial length of the photon pulse. Subsequently, each of the two excited qubits emits a real photon which is then captured by a local single-photon counting detector~\cite{Hadfield2009NPhoton}, leading to two remote measurement clicks. In this way, TRESP is manifested in the simultaneous excitation of the two remote qubits by a single high-frequency photon and the two subsequent low-frequency single-photon counting counts. The nonlocal action of a single photon has been revealed in the fruitful investigations on the single-photon two-mode entanglement~\cite{Tan1991PRL,Hardy1994PRL,Enk2005PRA,Hessmo2004PRL,Salart2010PRL} and EPR steering based on the homodyne measurement~\cite{Fuwa2015NCom}. However, in these works, a single photon always leads to a single counting event, but is impossible to trigger two counts as has been well elaborated recently~\cite{Aiello2021arXiv,Aiello2023Quantum}. Our findings differ fundamentally from these studies by demonstrating the ability of a single photon to yield two remote single-photon counting counts. Therefore, TRESP described here justifies Einstein's thought experiment from a completely new and fundamental perspective.

In addition, our model assumes that each of the two qubits is coupled to a single-mode cavity in the ultra-strong coupling~(USC) regime~\cite{Ciuti2006PRA, FornDiaz2019RMP,Kockum2019NatRevPhys}. It has been recently demonstrated that in this USC regime two qubits placed in a single-mode cavity can be excited simultaneously by a single  photon~\cite{Garziano2016PRL,Tomonaga2023arXiv}, which, however, can not be considered as a manifestation of TRESP. This is because that the single-mode cavity does not taken into account the spatial degree of freedom of the photon, and thus the whole cavity mode volume has to be considered as a single place for the photon detection. Furthermore, it has been shown that a single photon can excite two qubits individually placed in two coherently coupled cavities~\cite{Garziano2020SciRep}, which, again, can not be considered as TRESP since no spatial distance between the two cavities is included. 

Our model is based on the well-known cascade quantum system~\cite{Kolobov1987,Carmichael1993PRL,Gardiner1993PRL} which has served well the purpose of achieving quantum state transfer and remote entanglement~\cite{Cirac1997PRL,Muschik2013PRL,Vermersch2017PRL,Xiang2017PRX,Ritter2012Nature,Axline2018NPhys,Kurpiers2018Nature}, as shown in Fig.~\ref{fig1}(b). It consists of two qubit-cavity subsystems which are unidirectionally and dissipatively coupled from the left to the right with coupling strength determined by the cavity decay rate $\kappa_{j}$. For each subsystem, the interaction between the two-state (denoted by the ground state $|g\rangle$ and the excited state $|e\rangle$) qubit and the cavity is considered in the USC regime, namely, the coupling strength $g_{j}$ is comparable to the qubit transition frequency $\omega_{qj}$. USC between light and matter has been reached in several systems including superconducting quantum circuits~\cite{Niemczyk2010NPhys,FornDiaz2010PRL,Yoshihara2017NPhys}, optomechanics~\cite{Benz2016Science}, intersubband polaritons~\cite{Ciuti2005PRB,Anappara2009PRB,Gunter2009}, organic molecules~\cite{Gambino2014ACSPhoton} and Landau polaritons~\cite{Bayer2017NanoLett}, and has been an emerging frontier to explore new physics beyond the standard quantum Rabi model~\cite{Xie2017JPA}. Under the Coulomb gauge, we obtain the gauge-invariant Hamiltonian for each subsystem~(see Supplementary
Section I)
\begin{align}
\hat{\mathcal{H}}_{j}/\hbar =& \omega_{cj}\hat{a}^{\dagger}_{j}\hat{a}_{j}+ \frac{\omega_{qj}}{2}\big\{\hat{\sigma}_{jz}[\sin^{2}\theta_{j} + \cos^{2}\theta_{j}\cos(2\eta_{j}\hat{\mathcal{X}}_{j})] \nonumber\\[2ex]
+& \hat{\sigma}_{jy}\cos\theta_{j}\sin(2\eta_{j}\hat{\mathcal{X}}_{j})+\hat{\sigma}_{jx}\sin(2\theta_{j})\sin^{2}(\eta_{j}\hat{\mathcal{X}}_{j})\big\}
\label{Original-Hamiltonian}
\end{align} 
where $\omega_{cj}$ and $\hat{a}_{j}$ are the cavity resonance frequency and annihilation operator respectively, and $\hat{\mathcal{X}}_{j}=\hat{a}^{\dagger}_{j}+\hat{a}_{j}$.  $\hat{\sigma}_{jx}=\hat{\sigma}_{j} +\hat{\sigma}^{\dagger}_{j},\hat{\sigma}_{jy}=i(\hat{\sigma}_{j} -\hat{\sigma}^{\dagger}_{j})$ and $\hat{\sigma}_{jz} =\hat{\sigma}^{\dagger}_{j}\hat{\sigma}_{j} -\hat{\sigma}_{j}\hat{\sigma}^{\dagger}_{j}$ with $\hat{\sigma}_{j}=|g_{j}\rangle\langle  e_{j}|$ being the lowering operator of the $j$th qubit. $\eta_{j} = g_{j}/\omega_{qj}$ is the dimensionless coupling strength. $\tan\theta_{j} = (d^{(j)}_{ee} - d^{(j)}_{gg})/(2d^{(j)}_{eg})$ with $\{d^{(j)}_{ee}, d^{(j)}_{gg}\}$ and $d^{(j)}_{eg}$ denoting the permanent and transition dipole moments respectively. It has been pointed out that the permanent dipole moment is essential to induce parity-breaking transitions which are required to allow a single photon exciting two independent atoms~\cite{Garziano2016PRL}. We note that the form of $\mathcal{H}_{j}$ differs from that given in Ref.~\cite{Garziano2016PRL} which is not gauge-invariant. In the absence of the permanent dipole, i.e., $\theta_{j}=0$, Eq.~(\ref{Original-Hamiltonian}) recovers the recent results obtained in Ref.~\cite{DiStefano2019NPhys}, and reduces to the standard quantum Rabi model in the limit of $\eta_{j}\ll1$. 

Due to the inclusion of the counter-rotating terms, it is well known that the ground state of $\hat{\mathcal{H}}_{j}$, denoted by $|\emptyset_{j}\rangle$, contains a finite population of virtual excitation in the bare-state basis which can not be experimentally detected~\cite{Ridolfo2012PRL,Garziano2013PRA,Stassi2013PRL}. This would lead to unphysical nonzero cavity output and qubit excitation for $|\emptyset_{j}\rangle$ if observables are expressed in the bare-state basis. In order to correctly calculate the  physical observables, the system operators have to be expressed in the eigenstate basis of $\hat{\mathcal{H}}_{j}$~\cite{Ridolfo2012PRL,Garziano2013PRA}, i.e.,~$\hat{A}_{j}=\sum_{m_{j},n_{j}>m_{j}}\langle m_{j}|\hat{\mathcal{X}}_{j}| n_{j} \rangle |m_{j}\rangle \langle n_{j}|$ and $\hat{S}_{j}=\sum_{m_{j},n_{j}>m_{j}}\langle m_{j}|\hat{\sigma}_{jx}\cos\theta_{j}+\hat{\sigma}_{jz}\sin\theta_{j}| n_{j} \rangle |m_{j}\rangle \langle n_{j}|$ such that $\hat{A}^{\dagger}_{j}\hat{A}_{j}$ and $\hat{S}^{\dagger}_{j}\hat{S}_{j}$ represent the actual cavity and qubit excitation respectively~(see Supplementary
Section II).  For $\theta_{j}=0$, $\hat{S}_{j}$ recovers the expression commonly seen in the parity-preserving cases~\cite{Ridolfo2012PRL,Garziano2013PRA}. Here $\hat{\mathcal{H}}_{j}|n_{j}\rangle = \hbar\omega_{jn}|n_{j}\rangle$ with the eigenstate $|n_{j}\rangle$ arranged in an increasing order with respect to the eigenenergy $\hbar\omega_{jn}$. For clarity, we will use capital and calligraphic symbols to represent the operators in the eigenstate and bare-state basis respectively. 

Furthermore, each cavity is weakly coupled to an unidirectional propagating field $\hat{\mathcal{E}}(x,t)$. The input-output relation can then be derived as $\hat{\mathcal{E}}_{\text{out}}(x_{j},t) = \hat{\mathcal{E}}_{\text{in}}(x_{j},t) + \sqrt{\kappa_{j}/c}\hat{A}_{j}(t)$~\cite{Ridolfo2012PRL,Garziano2013PRA}, where $x_{j}$ denotes the position of the $j$th cavity respectively, and $c$ is the speed of light in vacuum~(see Supplementary
Section III). $x_{2} -x_{1} = d$ is the distance between the two cavities.
We thus have $\hat{\mathcal{E}}_{\text{in}}(x_{2},t) = \hat{\mathcal{E}}_{\text{out}}(x_{1},t-d/c) = \hat{\mathcal{E}}_{\text{in}}(x_{1},t-d/c) + \sqrt{\kappa_{1}/c}\hat{A}_{1}(t-d/c)$, in which $d/c$ is the time delay between the two spatially separated subsystems. From now on we will consider a narrow-band single-photon pulse which is an elementary excitation of the electromagnetic field defined as $|1\rangle=\int^{\infty}_{-\infty}dx u(x)\hat{\mathcal{E}}^{\dagger}_{\text{in}}(x)|\text{vac}\rangle$ with $u(x)$ being the spatial mode function satisfying the normalization relation $\int^{\infty}_{-\infty}dx |u(x)|^2=1$. The single-photon pulse is incident on the two subsystems which are in their respective ground state $|\emptyset_{j}\rangle$. In correspondence to Einstein's thought experiment, the single-photon pulse with finite spatial length plays the role of the diffracted single quantum particle and the two subsystems serve as two spatially separated ``imaginary'' detectors to measure the nonlocal action at two remote places. The spontaneously emitted photon from the $j$th qubit with decay rate $\gamma_{j}$ can then be collected and registered by the real single-photon detector $D_{j}$, otherwise an additional cavity can be employed to weakly couple to the qubit in order to more efficiently read out the qubit excitation~\cite{Kurpiers2018Nature}. One can then infer TRESP from the joint excitation of the two qubits, which is characterized by the second-order qubit correlation $C(t)=\langle\hat{S}^{\dagger}_{2}(t)\hat{S}^{\dagger}_{1}(t)\hat{S}_{1}(t)\hat{S}_{2}(t)\rangle$. If $C(t)\neq 0$ for $d\neq0$, the two single-photon detectors $D_{1}$ and $D_{2}$ click simultaneously, we can then conclude that a single photon has caused two counting events at two remote places. 

The local dynamics of each subsystem is not affected by the time delay $d/c$ due to the unidirectional coupling~\cite{Gardiner1993PRL,Cirac1997PRL},
however, the quantum correlation between the two subsystems, for instance, $C(t)$, would crucially depend on $d/c$. One would expect that $C(t)\rightarrow0$ for $d/c\rightarrow\infty$ since the single-photon pulse becomes increasingly unable to interact with the two subsystems simultaneously. 
In order to calculate the system dynamics, we first define the operators in the advanced time frame $\hat{A}^{'}_{2}(t)=\hat{A}_{2}(t+d/c)$ and $\hat{S}^{'}_{2}(t)=\hat{S}_{2}(t+d/c)$ for the second subsystem. We can then simplify the problem by integrating out the infinite degree of freedom for the propagating field using the Ito approach~\cite{Gardiner1985PRA,Combes2017AdvPhys}, and obtain a set of coupled master equations~(see Supplementary
Section IV)
\begin{align}
 \frac{d\rho_{\alpha\beta}(t)}{dt} =& \mathcal{L}\rho_{\alpha\beta}(t) + \sqrt{\alpha c}u(x_{1}-ct)[\rho_{\alpha-1,\beta}(t), \hat{L}^{\dagger}_{0}]\nonumber\\[2ex]
  &+ \sqrt{\beta c}u^{*}(x_{1}-ct)[\hat{L}_{0}, \rho_{\alpha,\beta-1}(t)]
  \label{master-eq}
\end{align}
where $\mathcal{L}\rho_{\alpha\beta}(t) = -i\hbar^{-1}[\hat{H}, \rho_{\alpha\beta}(t)]+ \sum_{n}\hat{L}_{n}\rho_{\alpha\beta}(t)\hat{L}^{\dagger}_{n}-\frac{1}{2}\{\hat{L}^{\dagger}_{n}\hat{L}_{n},\rho_{\alpha\beta}(t)\}$, $\rho_{\alpha\beta}(t)$ is the reduced density matrix for the coupled qubit-cavity system when the density matrix of the incident single-photon pulse is $|\alpha\rangle\langle \beta|$ ($\alpha,\beta\in\{0,1\}$). $\hat{H} = \hat{H}_{1} + \hat{H}_{2} + \frac{i\hbar}{2}\sqrt{G\kappa_{1}\kappa_{2}}(\hat{A}^{\dagger}_{1}\hat{A}^{'}_{2} - H.c.)$ which takes the typical form for a cascade system~\cite{Kolobov1987,Carmichael1993PRL,Gardiner1993PRL}, and $\hat{H}_{j}=\sum_{n}\hbar\omega_{jn}|n_{j}\rangle\langle n_{j}|$. Note that the dissipative coupling terms between the two cavities appearing in $\hat{H}$, i.e., $\frac{i\hbar}{2}\sqrt{G\kappa_{1}\kappa_{2}}(\hat{A}^{\dagger}_{1}\hat{A}^{'}_{2} - H.c.)$ are not expressed in terms of the bare-cavity mode operator but rather in the eigenstate basis of $\hat{H}_{j}$. Eq.~(\ref{master-eq}) shares the same form as previously presented in Refs.~\cite{Gheri1998FdP,Baragiola2012PRA}. The dissipative channels are $\{\hat{L}_{n}\}=\{\sqrt{\kappa_{1}}\hat{A}_{1}+\sqrt{G\kappa_{2}}\hat{A}^{'}_{2},\sqrt{\kappa_{2}(1-G)}\hat{A}^{'}_{2},\sqrt{\gamma_{1}}\hat{S}_{1},\sqrt{\gamma_{2}}\hat{S}_{2}\}$.  $G$ is the factor accounting for the possible power loss for the photon propagating from $x_{1}$ to $x_{2}$,  smaller $G$ indicates stronger photon loss. In the advanced time frame, the equal-time correlation is changed to a two-time correlation as $C(t)=  \langle\hat{S}^{'\dagger}_{2}(t-d/c)\hat{S}^{\dagger}_{1}(t)\hat{S}_{1}(t)\hat{S}^{'}_{2}(t-d/c)\rangle$, whose expectation value can be calculated based on the quantum regression theorem~\cite{Gheri1998FdP,Baragiola2012PRA}.

\begin{figure}[t!]
	\centering
	\includegraphics[width=0.49\textwidth]{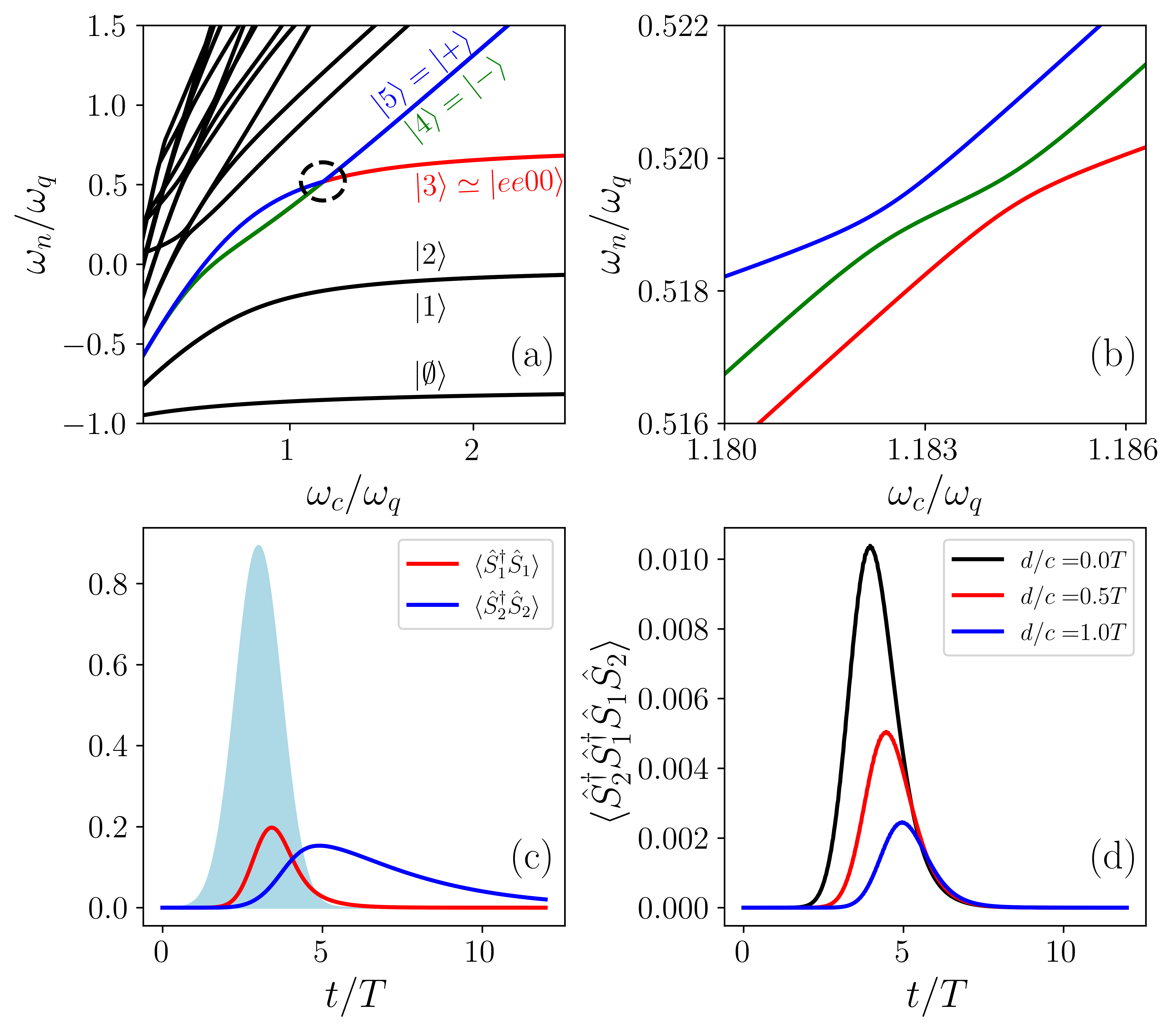}
	\caption{(a) The eigenenergy spectrum of $\hat{H}$ as a function of $\omega_{c}/\omega_{q}$. The first six eigenstates are labeled by $|n\rangle$, and $|\pm\rangle\simeq (|gg10\rangle \pm |gg01)/\sqrt{2}$ for large $\omega_{c}/\omega_{q}$.  (b) An enlarged view of the region denoted by the dashed circle in (a). There are two avoided crossings which indicates direct coupling between the states $|3\rangle$ and $|4\rangle$, and $|4\rangle$ and $|5\rangle$. (c) plots the excitation dynamics of each qubit, while the shaded region indicates the dimensionless mode function of the input photon pulse at the first cavity, i.e., $\sqrt{cT}u(x_{1}-ct)$. Here we have set $d=0$. (d) shows the second-order correlation between the two qubits for three different $d$. Here we have chosen parameters for the two subsystems as $\omega_{q1}=\omega_{q2}=\omega_{q}, \omega_{c1}=\omega_{c2}=\omega_{c},\theta_{1}=\theta_{2}=\theta,\eta_{1}=\eta_{2}=\eta, \gamma_{1}=\gamma_{2}=\gamma$. And $\eta = 0.5,\theta=\pi/5, G= 1.0, \kappa_{1}/\omega_{q} = 0.004,\kappa_{2}/\omega_{q}=0.001,T=1500/\omega_{q}$.
	}
	\label{fig2}
\end{figure}

\begin{figure*}[t!]
	\centering
	\includegraphics[width=0.95\textwidth]{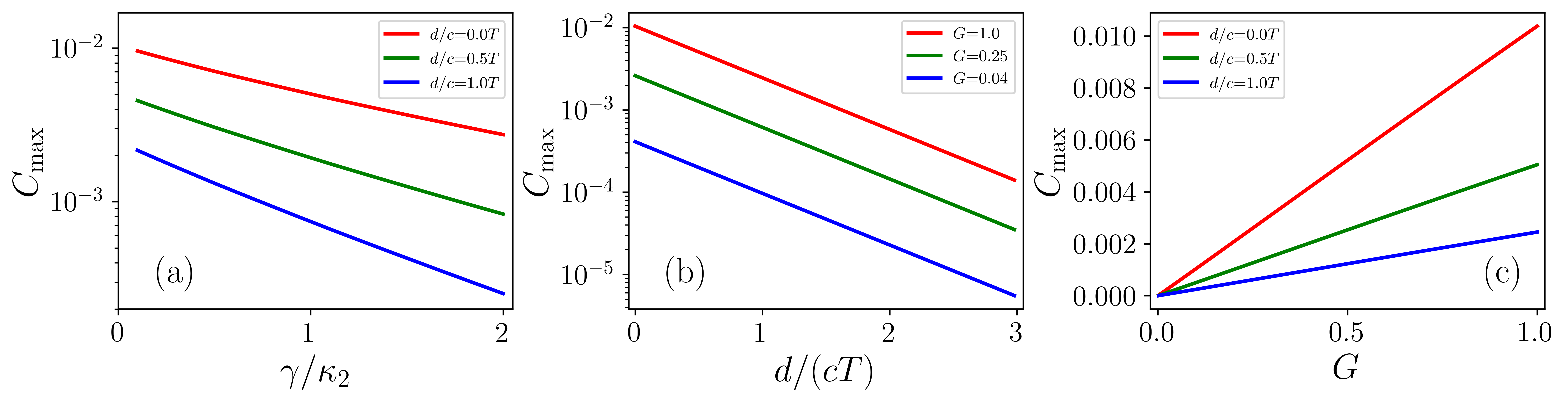}
	\caption{
		(a) $C_{\text{max}}$ as a function of $\gamma$ for $d=0$ and $G=1$. (b) and (c) plot $C_{\text{max}}$ as a function of the distance $d$ and power gain factor $G$ respectively for $\gamma=0$. Other parameters are the same as in the caption of Fig.~\ref{fig2}.
	}
	\label{parameter_dependence}
\end{figure*}

Having obtained the master equation for the composite system, our first step is to analyze the energy spectrum of $\hat{H}$. Considering the complicate structure of $\hat{\mathcal{H}}_{j}$, analytic analysis is almost impossible. We then turn to numerically calculate the energy spectrum as a function of $\omega_{c}/\omega_{q}$, which is shown in Fig.~\ref{fig2}(a) and (b), where $\omega_{n}$ is defined as $\hat{H}|n\rangle = \hbar\omega_{n}|n\rangle$ with $|n\rangle$ being the corresponding eigenstate and is ordered as $\omega_{n}\geq\omega_{m}$ for $n>m$. Here we choose the cavity decay $\kappa_{j}$ to be much smaller than $\omega_{q}$ in accordance with the assumption of weak cavity-field coupling. By choosing a relatively large $\eta = 0.5$, it is clear that there are two avoided crossings between the three eigenstates $|3\rangle, |4\rangle$ and $|5\rangle$, which in the limit of $\omega_{c}/\omega_{q}\gg 1$ can be written as $|3\rangle \simeq |ee00\rangle,|4\rangle\simeq (|gg10\rangle + |gg01\rangle)/\sqrt{2},|5\rangle\simeq (|gg10\rangle -|gg01\rangle)/\sqrt{2}$. Here $|\alpha\beta ij\rangle$ is a product state of the composite system expressed in the bare-state basis where the two qubits are in state $|\alpha\rangle$ and $|\beta\rangle$ and the two cavities are in Fock state $|i\rangle$ and $|j\rangle$ respectively, with $\alpha,\beta\in\{g,e\}$ and the integers $i,j\geq 0$. These two-excitation avoided crossings are formed through virtual transition processes due to the counter-rotating terms~\cite{Garziano2016PRL,Garziano2020SciRep}.
Around these avoided crossings, there are other non-negligible bare-state components contained in these eigenstates, nevertheless, it clearly indicates effective couplings between the doubly-excited qubit states and single-photon states. 
Though these avoided crossings are pretty small, the corresponding coupling strength is comparable to $\kappa_{j}$. We thus expect that the single-photon incident on the two coupled cavities would enable the joint excitation of the two spatially distant qubits, in other words, the single photon pulse can produce a nolocal action at two remote places.
Furthermore, it is worthy mentioning that these two-excitation crossings sit on the upper branches of the single-excitation crossings $\{|1\rangle,|2\rangle\}\leftrightarrow\{|4\rangle,|5\rangle\}$ for which the effective coupling strength are orders-of-magnitude stronger. Here $|1\rangle\simeq(|ge00\rangle - |eg00\rangle)/\sqrt{2}$ and $|2\rangle\simeq(|ge00\rangle + |eg00\rangle)/\sqrt{2}$ for $\omega_{c}/\omega_{q}\gg1$.

When the composite system is in its ground state, i.e., $|\emptyset\rangle = |\emptyset_{1}\emptyset_{2}\rangle$, the coupling terms in $\hat{H}$, i.e., $\frac{i}{2}\sqrt{G\kappa_{1}\kappa_{2}}(\hat{A}^{\dagger}_{1}\hat{A}^{'}_{2} - H.c.)$, can not cause any temporal dynamics in the absence of the incident photon. Thus, all observables expressed in terms of $\hat{A}_{j}$ or $\hat{S}_{j}$ have zero expectation values for $|\emptyset\rangle$, e.g., $C(t)=0$.
We then consider an incident single-photon  pulse in the spatial mode  $u(x)=\mathcal{N}\Theta(x_{1}-x)e^{-(x-x_{0})^2/(cT)^2}e^{i\omega_{\text{in}}x/c}$ with $\mathcal{N}$ being the normalization coefficient. $\Theta(x)$ is the step function defining the starting point of the pulse. $T$ is the temporal duration of the pulse and is chosen large enough comparing to $\kappa_{j}$ such that it is the longest timescale of the composite system. The central frequency of the photon pulse is chosen as $\omega_{\text{in}} = (\omega_{4}+\omega_{5} -2\omega_{0})/2$ which lies in the middle of the two-excitation avoided crossing between $|4\rangle$ and $|5\rangle$. We further take $\omega_{c}/\omega_{q}$ at the point when $\omega_{5}-\omega_{4}$ is minimized to have strongest coupling for the transition $|4\rangle\leftrightarrow|5\rangle$. Alternatively, it is equally possible to choose the other avoided crossing between $|3\rangle$ and $|4\rangle$. After numerically solving Eq.~(\ref{master-eq}), the temporal dynamics is plotted in Fig.~\ref{fig2}(c) and (d). 
For $d=0$,  the single excitation 
$\langle\hat{S}^{\dagger}_{2}\hat{S}_{2}\rangle$ is delayed with respect to $\langle\hat{S}^{\dagger}_{1}\hat{S}_{1}\rangle$ owing to the interaction between the photon and the first cavity as shown in Fig.~\ref{fig2}(c). Nonzero $d/c$ will only shift $\langle\hat{S}^{\dagger}_{2}\hat{S}_{2}\rangle$ in time as $t\rightarrow t+d/c$. In contrast, both the peak and temporal width of $C(t)$ decreases for larger $d/c$ as depicted in Fig.~\ref{fig2}(d). Nevertheless, $C(t)$ is obviously visible for a relatively large separation $d = cT$, thereby demonstrating the TRESP that the single-photon pulse has caused a simultaneous action at two remote places. Moreover, we notice that the TRESP has lead to a remote quantum entanglement between the two qubits, where they are either both in the excited state or both in the ground state.

The results discussed above are obtained in the optimal condition when the qubit decay and propagation loss are negligible, i.e., $\gamma_{1}=\gamma_{2}=0$ and $G=1$. Now we would like to include the effects of these dissipation channels. In Fig.~\ref{parameter_dependence}(a) we show the effect of the qubit decay on $C_{\text{max}}$ for $\gamma=\gamma_{1}=\gamma_{2}$, where $C_{\text{max}}=\text{Max}[C(t)]$. $C_{\text{max}}$ is not significantly affected by $\gamma$ even when $\gamma$ is comparable to the cavity decay $\kappa_{j}$, suggesting that the TRESP is robust against the qubit decay.
We next analyze the effect of the spatial separation $d$ on $C_{\text{max}}$, which is depicted in Fig.~\ref{parameter_dependence}(b) where $C_{\text{max}}$ as a function of $d/c$ is given for different $G$. $C_{\text{max}}$ decreases exponentially for increasing $d/(cT)$ for our choice of a Gaussian single-photon pulse. For $d=2cT$, $C_{\text{max}}$ is reduced roughly by a factor of 10. Considering a microsecond microwave photon pulse for the possible experimental implementation of our scheme based on superconducting quantum circuits, $d$ can be in the order of a few kilometers which is already in the macroscopic scale.  Finally, we consider the possible photon number loss denoted by $G$ which leads to two effects: Firstly, it reduces the photon number incident on the second cavity; Secondly, it shifts the two-excitation avoided crossings in the eigenenergy spectrum as indicated by the coupling terms in $\hat{H}$. Since $C(t)$ depends on the excited-state population of the second qubit which in turn is determined by the photon number entering the second cavity, $C_{\text{max}}$ then decreases linearly versus $G$ as can be seen from Fig.~\ref{parameter_dependence}(c).

We are now prepared to discuss the physical implication of our findings. First of all, our results support Einstein's thought experiment that a single quantum particle can produce an action at two or several places from a completely new perspective by demonstrating the ability of a single photon to yield two remote single-photon counting events simultaneously. Secondly, our findings sheds light on the fundamental aspects of quantum mechanics. 
For instance, our results indicates that the intuitive picture that a quantum particle can only give rise to a single detection count at one place is unnecessary. Moreover, it would be important to ask if TRESP is unique for photons or can be also found for other elementary particles such as electrons. 
%
TRESP demonstrated here has practical applications as well. Since the two qubits are entangled with each other, TRESP can be employed to create remote entanglement which is essential in building large-scale quantum networks~\cite{Ritter2012Nature,Axline2018NPhys,Kurpiers2018Nature}. The advantage of TRESP is that the frequency of the input photon differs significantly from that of the cavity, making it possible to create remote entanglement through an otherwise dissipative channel around the cavity resonance frequency.

In summary, we have developed a new model to demonstrate TRESP based on two unidirectionally coupled qubit-cavity subsystems in which the qubit-cavity interaction is assumed in the USC regime. We find that a single-photon pulse can excite simultaneously two remote qubits and give rise to two subsequent single-photon counting events even when the qubits are spatially separated by a distance comparable to the spatial length of the photon pulse. Considering that USC has been achieved in several physical systems, we would expect that our theoretical model is experimentally feasible. The required parity-breaking transitions can be realized in superconducting qubits~\cite{Niemczyk2010NPhys,LiuPRL2005}. In principle, TRESP illustrated here can be straightforwardly extended to a single particle detected at three or multiple places by adding more qubit-cavity subsystems to the cascade system and increasing the coupling strength to the deep-strong coupling regime~\cite{Bayer2017NanoLett}. 



The author is very grateful to the enlightening discussions with Klaus M{\o}lmer, Thomas Pohl and Tao Peng. The author would like to especially thank Thomas Pohl for his generous support and encouragement. This work was funded by the European Commission through the H2020-FETOPEN project ErBeStA (No. 800942), and the Center of Excellence ``CCQ'' (Grant Agreement No. DNRF152).

\bibliography{references}

\end{document}